\documentclass[12pt]{article}
\usepackage{a4,amsmath,amsxtra,amssymb}

\begin{document}

\baselineskip=1.5\baselineskip

\noindent{\textbf{\Large A  Twisting  Electrovac  Solution of Type
II with the Cosmological Constant }}

\vglue30pt

\noindent\textbf{J. K. Kowalczy\'nski}\footnote[1]{Institute of
Physics, Polish Academy of Sciences, Al. Lotnik\'ow 32/46, 02-668
Warsaw, Poland. E-mail: jkowal@ifpan.edu.pl}\hfill

\vglue60pt

\noindent An  exact  solution  of  the  current-free
Einstein--Maxwell equations  with  the  cosmological  constant  is
presented.  It is  of Petrov  type  II,  and  its  double
principal  null vector is geodesic,  shear-free,  expanding,  and
twisting.  The solution contains  five  constants.  Its
electromagnetic  field is  non-null and  aligned.  The  solution
admits only one Killing vector and
includes,  as  special cases,  several known  solutions.

\vskip30pt


\newpage

\noindent This  note  presents  an  exact  and  explicit  solution
of the current-free  Einstein--Maxwell  equations  with  the
cosmological constant.  The  solution  in  question  may  be
written  in  the  form
$$ds^{2} = 2\left(r^{2} +
n^{2}\right)d\zeta\,d\bar{\zeta} + 2\,dr\,k_{\mu}dx^{\mu} +
W\left(k_{\mu}dx^{\mu}\right)^{2}$$
with  the  electromagnetic
field  tensor
\begin{align*}
F_{\zeta\bar\zeta} = {}&\textstyle{1\over2}b\left(\zeta -
\bar{\zeta}\right) + in\left\{\zeta\left[a - {3\over2}b\left(r +
in\right)^{-1} - iA\right]\right.\\
&\textstyle{} +
\left.\bar{\zeta}\left[a - {3\over2}b\left(r - in\right)^{-1} +
iA\right]\right\}\,,\\
F_{\zeta u} = {}&-a + \textstyle
{1\over2}b\left(r + in\right)^{-1} + iA\,,\\
F_{\zeta r} =
in\bar{\zeta}F_{ur}\,,
\\
F_{ur} = {}&\textstyle {1\over2} b\left[\zeta\left(r +
in\right)^{-2} + \bar{\zeta}\left(r - in\right)^{-2}\right]\,,
\end{align*}
where $$k_{\mu}dx^{\mu} = du + in\left(\bar{\zeta}d\zeta - \zeta
d\bar{\zeta}\right)\,,$$
\begin{align*}
W := {}&\textstyle\left(r^{2} +
n^{2}\right)^{-1}\left[\Lambda\left({1\over3}r^{4} + 2n^{2}r^{2} -
n^{4}\right) + 2r\left(m + 2ab\zeta\bar{\zeta} + Bu\right) -
b^{2}\zeta\bar{\zeta}\right]\,,\\ A := {}&
\left(2n\right)^{-1}\left(b + C\right)\,,\quad B := n^{-2}b\left(b
+ C\right)\,,\quad C := \pm\left(b^{2} -
4a^{2}n^{2}\right)^{1/2}\,,
\end{align*}
and where $\zeta$ and $\bar{\zeta}$ are complex and conjugate
coordinates, $r$ and $u$ are real coordinates, $\Lambda$ is the
cosmological constant, $m$ is an arbitrary real constant, and $a$,
$b$, and $n$ are real constant arbitrary to a certain extent.
Relations involving $a$, $b$, and $n$ are discussed below.

Our solution is of Petrov  type  II  iff $b  \neq  0$.  Its  double
Debever--Penrose vector is just $k^{\mu}$ determined by the 1-form
$k_{\mu}dx^{\mu}$ given above, i.e. $k^{\mu} = \delta_{r}^{\mu}$.
$k^{\mu}$ is geodesic and shear-free. The rates of expansion
$\theta$ and of rotation $\omega$ of $k^{\mu}$ are given by the
following complex equation: $$\theta + i\omega = \left(r +
in\right)^{-1}\,.$$ Thus, for every $r \neq 0$ we have $\theta
\neq 0$, and $\omega = 0$ iff $n = 0$. $k^{\mu}$ is also a
principal null vector of our electromagnetic field
$\left(k_{[\mu}F_{\nu]\tau}k^{\tau} = 0\right)$, i.e. our case is
aligned. This field is non-null iff $b \neq 0$. Another
Debever--Penrose vector (single if type II, double if type D; for
subcases of Petrov
type D see below), say~$l^{\mu}$, is determined by
$l_{\mu}dx^{\mu} = dr + {1\over2}Wk_{\mu}dx^{\mu}$. Our solution
admits only one Killing vector, say~${\xi}^{\mu}$, such that
$${\xi}^{\zeta} = i\zeta\,,\qquad {\xi}^{r} = {\xi}^{u} = 0\,.$$

Our  solution includes,  as  special
cases, several known solutions. They can be obtained by eliminating
some of the constants, without making infinite values of course.
Note that $A$ and~$B$, and thus~$C$, must be real.

If  we  put  $a = b  = 0$,  then  we  eliminate  the  electromagnetic
field and obtain the well-known luxonic variant (zero Gaussian
curvature of a 2-space with the metric $\left(r^{2} +
n^{2}\right)d\zeta\, d\bar{\zeta}$, $r = {\rm constant}$) of the
Taub--NUT solution with the cosmological constant. This solution,
found by many authors, is of Petrov type D iff $m \neq 0$ or $n\Lambda
\neq 0$.

If  we  want  to  obtain  subsolutions  with  the  electromagnetic
field  but  without  the  rotation  $(n  =  0)$,  then  we have to
assume that  $a \neq 0$ or $b \neq 0$. If we put $b = 0$, then,
according  to  our assumption, we have  to keep $a \neq 0$. Then,
however, $A$ becomes imaginary, which  is forbidden. ($A$ occurs
as an additive term  in some  of $F_{\mu\nu}$'s expressed in terms
of only real coordinates,  e.g. when $\zeta = x  + iy$.) Thus  we
have  to assume that $b \neq 0$ (but only at the beginning of the
procedure, see below), and therefore we may not simply put $n = 0$
because  of the negative powers of $n$ in $A$ and~$B$. We may,
however, consider the limiting transition $n \to 0$.

If  $bC > 0$ ($C$ being  real  of  course),  then  the  limiting
transition  $n \to  0$ is  forbidden  since it  would make
infinities.

If  $bC < 0$ and $n \to 0$, then $A \to 0$, $B \to 2a^{2}$, and
our $ds^{2}$ falls under a category  of  metric  forms  for  which
all  the possible electromagnetic  fields  were
found~[1];\footnote[2]{In [1] the  signs of the
cosmological  constant  (denoted therein  by~$\lambda$) are
opposite  to  those  commonly assumed,  i.e. $\lambda =
-\Lambda$.} and then  we  obtain the solution (3.4) from~[1],
found earlier by Leroy~[2].\footnote[3]{In [2] this solution
is presented in a different coordinate system by eqs.~(6.12c).
It is quoted in the monograph [3] as eqs. (24.54d) where, in the
second equation, $ x$ should read $ {\rm e}^{x} $ (multiplied by a
proper constant; notation after~[3]). This misprint is corrected
on p.~11 in~[4], but the correction is there unfortunately related
to the next eqs. (24.54e) in~[3].} In this
solution, being of Petrov type II iff  $b \neq 0$, $a$ and $b$ are
independent. If  we  put $b = 0$, then  we obtain a
special case  of some  of the solutions
listed in~[1]. This special case ($ b = n = 0$) is of Petrov
type D iff $a \neq 0$ or $m \neq 0$, conformally flat iff
$a = m = 0$ and $\Lambda \neq 0$, and flat iff $a = m = \Lambda = 0$.

If we assume that $C = 0$, then our solution is still of Petrov type
II  and  twisting (iff $an \neq 0$,  since $b^{2} = 4a^{2}n^{2}$
in  this  case),  but it  contains  only  one electromagnetic
constant, $a$, and  does  not contain the negative  powers of $n$
in $A$ and~$B$. If we  put $n = 0$, then we obtain the
special  case described  at the end of the preceding paragraph.

The solution presented in this note should be considered as new
since, as far as I know, no solutions generalizing those listed
in~[1] (excluding solution (3.2) therein) have been published.

\vskip30pt


\noindent{\bf\large REFERENCES}

\vglue12pt

\noindent 1.  Bajer, K., and Kowalczy\'nski, J. K. (1985). {\it J.
Math. Phys.} {\bf26}, 1330.

\noindent 2.  Leroy,  J.  (1976).  {\it Bull. Cl. Sci. Acad. R.
Belg.} {\bf62},  259.

\noindent 3.  Kramer, D., Stephani, H., MacCallum, M. A. H., and
Herlt, E. (1980). {\it Exact Solutions of Einstein's Field Equations}
(VEB Deutscher Verlag der Wissenschaften, Berlin / Cambridge
University Press, Cambridge).

\noindent 4.  Kramer, D., Stephani, H., MacCallum, M. A. H., and
Herlt, E. (1984). ``Exact Solutions of Einstein's Field Equations:
Corrections." Preprint.


\end{document}